\documentstyle[12pt,epsfig]{article}
\def\be{\begin{equation}} 
\def\ee{\end{equation}} 
\def\bea{\begin{eqnarray}}
\def\eea{\end{eqnarray}} 
\def\nnb{\nonumber}
\def\dR{${\hat{d}}_R$}
\def\G3H{$\Gamma^{\beta}_t$}

\def\signud{{$\sigma_{\nu d}$}}

\begin{document}
\renewcommand{\thefootnote}{\fnsymbol{footnote}}
\setcounter{footnote}{1}

\hfill{December 12, 2002}

\hfill{\tt USC(NT)-02-2, SNUTP 02-014, TRI-PP-02-07}

\begin{center}
{\Large{\bf 
Solar-neutrino reactions on deuteron \\
\vskip 1mm
in effective field theory
}}
\vskip 5mm
{\large
S. Ando$^{(a)}$\footnote{E-mail address : 
sando@nuc003.psc.sc.edu},
Y. H. Song$^{(b)}$\footnote{E-mail address : 
singer@phya.snu.ac.kr},
T.-S. Park$^{(a)}$\footnote{E-mail address : 
tspark@nuc003.psc.sc.edu},
H. W. Fearing$^{(c)}$\footnote{E-mail address : 
fearing@triumf.ca},
\\ \vskip 1mm
and K. Kubodera$^{(a)}$\footnote{E-mail address : 
kubodera@sc.edu}}
\vskip 3mm
{\it
$^{(a)}$ Department of Physics and Astronomy, \\
University of South Carolina, Columbia, 
SC 29208, U.S.A.
\vskip 1mm

$^{(b)}$ Department of Physics, 
Seoul National University, 
Seoul 151-742, Korea
\vskip 1mm

$^{(c)}$ Theory Group, TRIUMF, Vancouver, 
BC V6T 2A3, Canada
\vskip 1mm
}
\end{center}

\vskip 5mm

The cross sections for low-energy neutrino-deuteron
reactions are calculated within
heavy-baryon chiral perturbation theory 
employing cut-off regularization scheme. 
The transition operators are derived
up to next-to-next-to-next-to-leading order 
in the Weinberg counting rules,
while the nuclear matrix elements are evaluated
using the wave functions generated by
a high-quality phenomenological NN potential. 
With the adoption of  
the axial-current-four-nucleon 
coupling constant fixed from
the tritium beta decay data, 
our calculation is free from unknown 
low-energy constants.
Our results exhibit a high degree of stability
against different choices of the cutoff parameter,
a feature which indicates that,
apart from radiative corrections,
the uncertainties in the calculated cross sections
are less than 1 \%.

\vskip 3mm


\newpage
\renewcommand{\thefootnote}{\arabic{footnote}}
\setcounter{footnote}{0}

\noindent
{\bf 1. Introduction} %

This Letter is concerned with a theoretical estimation
of the cross sections, \signud, for
the neutrino-deuteron reactions
\bea
&& \nu _e + d \to e^- + p + p, \ \ \ 
   \bar{\nu}_e + d \to e^+ + n + n \ \ \ \mbox{(CC)},
   \label{eq:CC}
\\ &&
     \nu _l + d \to \nu_l + p + n, \ \ \ \ 
   \bar{\nu}_l + d \to \bar{\nu}_l + p + n \ \ \ \ \, 
   \mbox{(NC)},
   \label{eq:NC}
\eea
where CC and NC stand for the charged-current and 
neutral-current reaction, respectively,
and $l$ denotes the lepton flavor ($l=e, \mu, \tau$).
Recent SNO experiments \cite{SNO,SNO2}
have provided strong evidence for $\nu_e$ oscillations.
In interpreting the existing and future
SNO data, accurate estimates of \signud\ 
in the solar neutrino energy region 
($E_\nu\le$ 20 MeV) are of great importance.

Recently,
two theoretical approaches have been used 
for evaluating \signud.
One is a traditional method in which
nuclear electroweak processes are described
in terms of one-body impulse approximation 
(IA) operators and two-body exchange-current 
(EXC) operators acting on non-relativistic 
nuclear wave functions.
The EXC contributions are derived 
from one-boson exchange diagrams \cite{cr},
while the nuclear wave functions are obtained
by solving the Schr\"{o}dinger equation
involving high-quality realistic nuclear interactions.
For convenience, we refer to this method 
as the standard nuclear physics approach (SNPA).
The successful applications of SNPA are
well documented in the literature \cite{cs}.
A detailed calculation of \signud\ 
based on SNPA was carried out 
by Nakamura, Sato, Gudkov and Kubodera (NSGK)
\cite{NSGK},
and this calculation has recently been updated
by Nakamura {\it et al.}\ (NETAL)  
\cite{Netal}.\footnote{
For earlier calculations,
see, {\it e.g.,} \cite{YHH,KN}.}

The second approach is based 
on effective field theory (EFT),
which has been gaining ground 
as a new tool for describing
low-energy phenomena in few-nucleon systems
\cite{weinberg,epel,EFT}.
Butler, Chen, and Kong (BCK) \cite{BCK} applied EFT
to the $\nu d$ reactions,
using the regularization scheme called
the power divergence subtraction (PDS)
\cite{PDS}.
Their results agree with those of 
NSGK in the following context.
The EFT Lagrangian in PDS involves 
one unknown low-energy constant (LEC),
denoted by $L_{1A}$,
which represents the strength of
axial-current-four-nucleon contact coupling.
BCK adjusted $L_{1A}$ to optimize
fit to the \signud\ of NSGK
and found that,
after this adjustment,
the results of the EFT and SNPA calculations
agree with each other within 1\%
over the entire solar-$\nu$ energy region
for all of the four reactions 
in Eqs.(\ref{eq:CC}) and (\ref{eq:NC}).
Furthermore, the best-fit value of $L_{1A}$
was found to be of a reasonable magnitude
consistent with the ``{\it naturalness}" argument
\cite{BCK}.

The fact that the results of 
an {\it ab initio} EFT calculation
(with one free parameter fine-tuned) are
consistent with those of SNPA is considered to 
give strong support for the basic soundness of SNPA.
At the same time, it highlights 
the desirability of an EFT calculation of \signud\ 
free from an adjustable parameter.
In this Letter we describe an attempt 
toward such a goal.
We employ here a formalism recently developed 
in the studies of the solar {\it Hep} process 
and the solar {\it pp} fusion reaction \cite{PKMR,pphep}.
In this method, invoking heavy-baryon chiral 
perturbation theory (HB$\chi$PT),  
we construct transition operators
from irreducible diagrams according to
Weinberg's counting scheme \cite{weinberg};
the nuclear matrix elements are evaluated
by sandwiching the EFT-controlled transition operators
between the nuclear wave functions that have been 
obtained by solving the Schr\"{o}dinger equation
involving high-quality realistic nuclear interactions.
For convenience, we refer to this EFT-motivated approach
as EFT*.  
It is known \cite{PKMR} that, 
for the present purposes, 
it is sufficient to consider
up to next-to-next-to-next-to-leading order (N$^3$LO)
in HB$\chi$PT, 
and that to this order there is only one
unknown LEC, denoted by \dR\ in \cite{pphep}.
Like $L_{1A}$ in \cite{BCK}, \dR\  controls
the strength of the axial-current-four-nucleon contact
coupling and subsumes short-distance physics 
that has been integrated out.
An important point noticed in \cite{pphep}
is that, since the tritium $\beta$-decay rate \G3H
is also sensitive to \dR,
we can determine \dR\ from 
the well-known experimental value of \G3H.
Once $\hat{d}^R$ is determined,
we can make a parameter-free calculation
of \signud, and
the purpose of this communication
is to describe such a calculation.\footnote{
Similar parameter-free calculations have been 
carried out for the solar $pp$-fusion reaction 
and the solar {\it hep} process~\cite{pphep},
and for $\mu$-$d$ capture~\cite{mud}.}
We shall show that, apart from radiative corrections
for which we refer 
to the literature~\cite{beacom,RC_towner,kurylov},
\signud\ given here is reliable with $\sim$ 1 \% precision.
 
\vskip 3mm \noindent
{\bf 2. Calculational method}

For low-energy processes, we can work with
the current-current interaction:
\bea
H = \frac{G_F'}{\sqrt{2}}\int d^3\vec{x} \left[
V_{ud}
J^{(\mbox{\tiny{CC}})}_\mu(\vec{x}) 
l^{(\mbox{\tiny{CC}})\mu}(\vec{x}) + 
J^{(\mbox{\tiny{NC}})}_\mu(\vec{x}) 
l^{(\mbox{\tiny{NC}})\mu}(\vec{x}) \right] ,
\eea
where $G_F'= 1.1803\times 10^{-5}$ 
[GeV${}^{-2}$]\cite{hardy} is 
the weak coupling constant, and 
$V_{ud}= 0.9746$ is the K-M matrix element.
$G_F'$ includes the inner radiative correction:
${G_F'}^2 = G_F^2(1+\Delta_R^V)$,
where 
$G_F = 1.1166\times 10^{-5}$ [GeV$^{-2}$]
is the Fermi constant and 
$\Delta_R^V$ is the inner radiative 
correction~\cite{hardy}.\footnote{
For more detail discussion of the radiative correction,
see \cite{Netal}.} 
The CC- and NC-lepton currents,
$l^{(\mbox{\tiny{CC}})\mu}$ and
$l^{(\mbox{\tiny{NC}})\mu}$, are well known;
the CC- and NC hadronic currents, 
$J^{(\mbox{\tiny{CC}})}_\mu$ and
$J^{(\mbox{\tiny{NC}})}_\mu$,
are written as 
\bea
J^{(\mbox{\tiny{CC}})}_\mu(\vec{x})
 &=& V^\pm_\mu(\vec{x})
- A^\pm_\mu(\vec{x}) ,
\label{eq;CC}
\\
J^{(\mbox{\tiny{NC}})}_\mu(\vec{x}) 
&=& (1-2{\rm sin}^2 
\theta_W)V^0_\mu(\vec{x})
- A^0_\mu(\vec{x})
-2{\rm sin}^2\theta_W V^S_\mu(\vec{x}) ,
\label{eq;NC}
\eea
where $V_\mu$ and $A_\mu$ represent
the vector and axial current, respectively.
The superscripts, $\pm$ and 0, are the isospin indices
of the isovector current and $S$ denotes the isoscalar current;
$\theta_W$ is the Weinberg angle, 
${\rm sin}^2\theta_W$= 0.2312.

The $\nu$-$d$ reactions can lead to
various values of the relative orbital angular momentum, 
$L$, of the final two nucleons.  
We concentrate here, however, 
on the $L$=0 state (${}^1S_0$),
since it is this partial wave that involves
the $\hat{d}^R$ term
and since the contributions of higher partial waves
are well understood in terms of the one-body operators.
The contributions from $L\!\ge\!1$ are significant
in the upper part of 
the solar neutrino energy region,\footnote{
The $L\!\ge\!1$ contributions increase 
$\sigma_{\nu d}$ by $\sim$3.8 \%
at $E_\nu$= 20 MeV \cite{Netal,kn}.}  
but their uncertainty is small enough
to be ignored in the present context.

The one-body (1B) currents can be obtained 
from the phenomenological form factors
of the weak-nucleon current.\footnote{
The low-energy structure of the form factors
has been studied in detail within HB$\chi$PT~\cite{1b-chiral}.
At N$^2$LO, however,
we in principle need to consider 
off-shell form factors~\cite{offshell},
a feature that reflects arbitrariness
in choosing fields~\cite{nooffshell}.
The influence of off-shell terms, however,
should be small at low energies;
see, {\it e.g.}, Ref.\ \cite{schwamb}.
An EFT study of the one-body Gamow-Teller
matrix element in the two-nucleon system~\cite{PMKR-1B}
explicitly shows that the off-shell effects
are sufficiently small for our present purposes.}
The isovector vector and axial-vector currents
are given in momentum space as
\footnote{Since we consider the final ${}^1S_0$ state only,
there is no contribution from the isoscalar current.}
\bea
J_{V}^{a\mu}(q) &=& \bar{u}(p') \frac{\tau^a}{2}
\left[ g_V(q)\gamma^\mu + 
g_M(q)\frac{i\sigma^{\mu\nu}q_\nu}{2m_N}\right]
u(p) ,
\label{eq;vector}
\\
J_A^{a\mu}(q) &=& \bar{u}(p')\frac{\tau^a}{2}
\left[ g_A(q)\gamma^\mu \gamma_5 +
 g_P(q)\frac{q^\mu}{m_\mu}\gamma_5\right]
u(p) ,
\label{eq;axial}
\eea
where `$a$' is the isospin index,
$u(p)$ is the Dirac spinor for the nucleon,
and $m_\mu$ ($m_N$) is the muon (nucleon) mass;
$g_V(q)$, $g_M(q)$, $g_A(q)$, and $g_P(q)$ are
the vector, magnetic, axial-vector, 
pseudoscalar form factors, respectively.
It is known empirically 
that the first three form factors can be parametrized 
very well in the dipole form
with the use of effective radii, 
$r_V^2=0.59$, $r_M^2=0.80$ and $r_A^2=0.42$ fm$^2$,
for $g_V(q)$, $g_M(q)$ and $g_A(q)$, 
respectively\cite{mergell}. 
We are adopting here the usual normalization:
$g_V(0)=1$, $g_A(0)$=$g_A$= 1.267, 
and $g_M(0)$=$\kappa_V=3.706$.
Although $g_P(q)$ is not well known empirically, 
it is strongly constrained by chiral symmetry;
an HB$\chi$PT calculation up to NNLO 
\cite{1b-chiral} leads to
\bea
g_P(q) = -\frac{2m_\mu f_\pi g_{\pi N}}{q^2-m_\pi^2}
-\frac13 g_A m_\mu m_N r_A^2,
\eea
where 
$g_{\pi\!N}=13.5$. 
In fact, the contribution of the $g_P$ term
is tiny in our case. 
We apply a non-relativistic expansion
of the above expressions
and retain terms up to ${\cal{O}}(1/m_N^3)$ 
(corresponding to N$^3$LO of the chiral order); 
the details will be described elsewhere \cite{song}.

The two-body (2B) current operators
are derived from the chiral lagrangian ${\cal L}$,
which is expanded as 
${\cal L} = \sum_{\bar{\nu}}{\cal L}_{\bar{\nu}}
= {\cal L}_0 + {\cal L}_1 + \cdots$,
where ${\cal L}_0$ and ${\cal L}_1$ are LO and NLO lagrangians,
respectively.  Their explicit expressions are:
\bea
{\cal L}_0 &=& \bar{N}\left[iv\cdot D+2ig_AS\cdot\Delta\right] N
+ f_\pi^2{\rm Tr}\left(-\Delta\cdot\Delta+\frac{\chi_+}{4}\right) ,
\\
{\cal L}_1 &=& \frac{1}{2m_N} \bar{N}\left[
(v\cdot D)^2 -D^2 + 2g_A\{v\cdot \Delta,S\cdot D\}
-(8\hat{c}_2-g_A^2)(v\cdot \Delta)^2
\right. \nnb \\ && \left.
-8 \hat{c}_3\Delta\cdot\Delta
-(4\hat{c}_4+1) [S^\mu,S^\nu][\Delta_\mu,\Delta_\nu]
-2i(1+\kappa_V)[S^\mu,S^\nu] f_{\mu\nu}^+
\right] N
\nnb \\ && + \frac{g_A}{m_Nf_\pi^2}\left[
-4i\hat{d}_1\bar{N}S\cdot\Delta N\bar{N}N
+2i\hat{d}_2 \epsilon^{abc}
\epsilon_{\mu\nu\alpha\beta}v^\mu\Delta^{a,\nu}
\bar{N}S^\alpha\tau^bN\bar{N}S^\beta \tau^cN\right],
\eea
where $v^\mu$ is the velocity vector $v^\mu=(1,\vec{0})$ and
$S^\mu$ is the spin operator $2S^\mu=(0,\vec{\sigma})$.
The explicit expressions of the fields, $D_\mu$, $\Delta_\mu$,
$f_{\mu\nu}^+$, and $\chi_+$, are given in \cite{mud},
and $f_\pi$ is the pion decay constant.
The LEC's,  $\hat{c}_i$, have been
determined by Bernard {\it et al.}\ 
at the tree-level~\cite{bernard-LECs-cs-treelevel};\footnote{
The relation between our dimensionless LEC's, 
$\hat{c}_i$'s, and $c_i$'s used in the literature
is $\hat{c}_i= m_N c_i$.}
\bea
\hat{c}_2 =  1.67\pm 0.09, \ \ \
\hat{c}_3 = -3.66\pm 0.08, \ \ \
\hat{c}_4 =  2.11\pm 0.08 .
\eea
The LEC's of the contact terms $\hat{d}_{1,2}$ will be discussed 
later in the text.

We construct 2B transition operators
from 2B irreducible Feynman diagrams
up to N$^3$LO
in Weinberg's counting rule~\cite{weinberg}.
Since the tree-level 2B operators are
higher in chiral counting 
than the tree-level 1B operators by two orders,
we can limit ourselves to tree diagrams for the 2B operators.
In addition, since the $g_P$ term
is highly suppressed,
we do not consider it in the 2B operators.

\begin{figure}[h]
\begin{center}
\epsfig{file=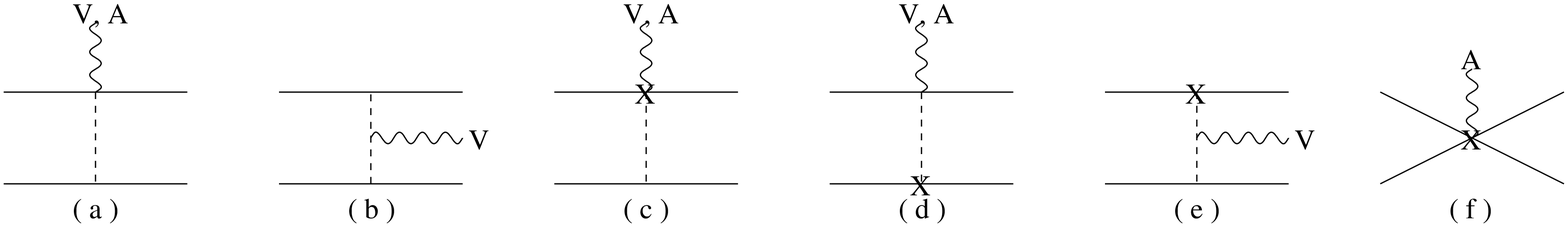,width=16cm}
\caption{Diagrams for two-body current operators 
of order $\nu=$ 1 (a,b) and $\nu$=2 (c,d,e,f).
The wavy lines with $V$ and $A$ attached
denote the vector and axial-vector current, respectively,
the dashed line denotes the pion, and
vertices without (with) ``X''
arise from the LO (NLO) lagrangian.
}
\label{fig;2Bcurrents}
\end{center}
\end{figure}
The diagrams for the 2B operators are 
given in Fig. \ref{fig;2Bcurrents}.
Since we only have nucleons and pions in ${\cal L}$,
the effects involving exchange of heavier mesons
such as the $\sigma$ and $\rho$ mesons 
are embedded in the contact term, 
diagram (f) in Fig. \ref{fig;2Bcurrents}.
We denote by $\Lambda$ a momentum scale
below which our nucleon-pion-only description
is expected to be valid.
To prevent the exchanged momentum 
from surpassing $\Lambda$, 
we introduce the cutoff function
$S_\Lambda(\vec{k}) = e^{-\vec{k}^2/(2\Lambda^2)}$
in calculating the Fourier transforms 
of the 2B transition operators \cite{pphep}.
As noted in \cite{pphep},
the short-range part of the 2B contributions
can be lumped together into an axial-current-four-nucleon
contact coupling term with the strength $\hat{d}^R$,
where
$\hat{d}^R = \hat{d}_1 + 2 \hat{d}_2 + 
\frac13\hat{c}_3 +\frac23 \hat{c}_4 + \frac16$.
Then, for a given value of $\Lambda$,
we can determine $\hat{d}^R$ from
the empirical value of \G3H.
The results are \cite{pphep}:
\bea
\hat{d}^R = 1.00 \pm 0.07, \ \  1.78\pm 0.08, \ \ 3.90\pm 0.10,
\label{eq;hatdR}
\eea
for $\Lambda$ = 500, 600, 800 MeV, respectively.
The explicit expressions of the current operators 
for the CC reaction
have been given in \cite{mud}.\footnote{
Insofar the final two-nucleon partial
wave is limited to $s$-wave,
one can use the same expression
for the NC reaction
(with an appropriate change 
in the coefficient of the vector current)}
\bea\nnb\eea
\vskip 3mm \noindent
{\bf 3. The total cross section}

The total cross sections are calculated using the
nonrelativistic formula:
\bea
\sigma_{\nu d}(E_\nu) &=& \int dp \int dy \frac{1}{(2\pi)^3}
\frac{2p^2 {k'}^2}{{k'}/E'+({k'}-E_\nu y)/(2m_N)}
F(Z,E') \frac{1}{3} \sum_{spin} |T|^2 ,
\eea
with the energy conservation relation valid up to $1/m_N$,
\bea
m_d+E_\nu-E'-2m_N
-\frac{1}{m_N}\left[p^2
+\frac14\left(E_\nu^2+{k'}^2-2E_\nu{k'}y\right)\right]=
 0 ,
\eea
where $E_\nu$ ($E'$) is the energy of 
the initial neutrino (final lepton),
$p$ is the magnitude of the relative three-momentum
between the final two nucleons,
$k'$ is that of the outgoing lepton ($k'=|\vec{k'}|$),
and $y$ is the cosine of the angle 
between the incoming and outgoing leptons
($y=\hat{k}_\nu\!\cdot\!\hat{k'}$).
$F(Z,E')$ is the Fermi function
and $m_d$ is the deuteron mass.
The transition matrix $T$ is decomposed as
$T=T_{1B}+T_{2B}$, where $T_{1B}$ and $T_{2B}$ 
are the contribution of the 1B and 2B operators, respectively.
These will be evaluated with the use of the 
Argonne V18 potential~\cite{av18}.

Since the calculation of $T_{1B}$ is standard
\cite{song},
we give here only the explicit expression for 
$T_{2B}$:
\bea
\lefteqn{\frac{1}{\sqrt{4\pi}}T_{2B} =
\beta \chi^\dagger_{00}\vec{\Sigma}\chi_{1m_d}
\int dr \left\{
\vec{F}_6 u_0(r) j_1(qr/2)
\frac{y_{1\Lambda}(r)}{r} u_d(r)
\right. } \nnb \\ &+&
\vec{F}_7 \left[
u_0'(r)  \left( u_d(r) -\sqrt{2}w_d(r)\right)
- u_0(r) \left( u_d'(r)-\sqrt{2}w_d'(r)\right) \right]
j_0(qr/2) \frac{y_{1\Lambda}(r)}{r}
\nnb \\ &+&
\vec{F}_8  u_0(r)j_0(qr/2)\frac{y_{1\Lambda}(r)}{r^2} w_d(r)
+ \vec{F}_9  u_0(r) j_0(qr/2) y_{0\Lambda}(r) u_d(r)
\nnb \\ &+&
 \vec{F}_{10}  u_0(r)j_0(qr/2)y_{2\Lambda}(r) w_d(r)
+\vec{F}_{11}  u_0(r) \delta_{\Lambda}(r) u_d(r)
\nnb \\ &+& \left.
\vec{F}_{12} \int^{1/2}_{-1/2} dy
 u_0(r) j_0(yqr)
\left[
y_{0\Lambda}^L(r) u_d(r)
-\frac23y_{1\Lambda}^L(r)
\left(u_d(r)+\frac{w_d(r)}{\sqrt{2}}\right)
\right]
\right\} ,
\label{eq;2Bamplitude}
\eea
where $\beta=(G_F'V_{ud})^2/2$ for CC 
and $\beta = {G_F'}^2/4$ for NC.
$\vec{\Sigma}=\vec{\sigma}_1-\vec{\sigma}_2$,
with $\vec{\sigma}_i$ being the $i$-th nucleon 
spin operator;
$\chi_{1m_d}$ and $\chi_{0,0}$ are the spin
wave functions for the deuteron
and the final two nucleons, respectively.    
The radial function $u_0$ corresponds 
to the final two-nucleon s-wave, 
while $u_d$ and $w_d$ are 
the s-wave and d-wave radial functions
of the deuteron;
$j_L(qr/2)$ is the spherical Bessel function;
$q^\mu$ is a momentum transfer between the currents,
$q^\mu={k'}^\mu-k^\mu$ and $q=|\vec{q}|$.
Furthermore, 
\bea
\vec{F}_6 &=&
-\frac{g_A}{f_\pi^2} v\cdot {\cal J} \hat{q}
-\frac{g_A(1+\kappa_V)}{2m_Nf_\pi^2} 
i\vec{q}\times(\hat{q}\times\vec{\cal J})
-\frac{g_A}{4 m_N f_\pi^2} \vec{q}\cdot\vec{\cal J}\hat{q},
\nnb \\
\vec{F}_7 &=& \frac{- g_A}{6m_Nf_\pi^2}\vec{\cal J} ,
\ \ \
\vec{F}_8 = \frac{- g_A}{\sqrt{2}m_Nf_\pi^2} \vec{\cal J},
\ \ \
\vec{F}_9 = - \frac{2g_A m_\pi^2\left(\hat{c}_3+2\hat{c}_4+\frac12\right)}
                   {3m_Nf_\pi^2} \vec{\cal J} ,
\\
\vec{F}_{10} &=&
\frac{2\sqrt{2}g_Am_\pi^2\left(\hat{c}_3-\hat{c}_4-\frac14\right)}
{3 m_Nf_\pi^2} \vec{\cal J},
\ \
\vec{F}_{11} = \frac{2g_A\hat{d}^R}{m_Nf_\pi^2} \vec{\cal J} ,
\ \
\vec{F}_{12} = \frac{-1}{2}
\left(\frac{g_A}{f_\pi}\right)^2 i(\vec{q}\times\vec{\cal J}) ,
\nnb
\eea
where ${\cal J}^\mu$ is the lepton current in momentum space, and
\bea
\delta_\Lambda(r)
= \int \frac{d^3\vec{k}}{(2\pi)^3}
e^{i\vec{k}\cdot\vec{r}}S_\Lambda^2(\vec{k}),
\ \ \ 
y_{0\Lambda}(r)
= \int \frac{d^3\vec{k}}{(2\pi)^3}
e^{i\vec{k}\cdot\vec{r}}
\frac{S_\Lambda^2(\vec{k})}{\vec{k}^2+m_\pi^2} , 
\label{eq;cutoff}
\eea
and $y_{1\Lambda}=-r\frac{d}{dr}y_{0\Lambda}(r)$,
$y_{2\Lambda}=\frac{r}{m_\pi^2}
\frac{d}{dr}\left[\frac{1}{r}\frac{d}{dr}
y_{0\Lambda}(r)\right]$;
$y_{0,1\Lambda}^L(r)$ is obtained
by exchanging the pion mass $m_\pi$
to $L = \sqrt{m_\pi^2+(1/4-y^2)\vec{q}^2}$
in Eq. (\ref{eq;cutoff}).
In the above expression we have neglected 
the small terms proportional to $q^2$.

\vskip 3mm \noindent
{\bf 4. Numerical results and discussion}

As mentioned, we consider in this work 
the contribution from the final 
two-nucleon s-wave only.  
The corresponding total cross section
is denoted by $\sigma_{\nu d}^{L=0}$.
Table \ref{table;cross-section} gives
$\sigma_{\nu d}^{L=0}$ calculated in EFT*
for the four reactions in Eqs.(1) and (2).
\begin{table}[h]
\begin{center}
\begin{tabular}{| c | l l l l |} \hline
$E_\nu$ & $\nu d \to e^-pp$ & $\bar{\nu}d\to e^+nn$ &
$\nu d \to \nu np$ & $\bar{\nu}d\to\bar{\nu}np$ \\ \hline
2       & 0.004    & 0        & 0        & 0       \\
3       & 0.047    & 0        & 0.003    & 0.003   \\
4       & 0.158    & 0        & 0.031    & 0.031   \\
5       & 0.348    & 0.029    & 0.096    & 0.094   \\
6       & 0.625    & 0.120    & 0.204    & 0.198   \\
7       & 0.996    & 0.284    & 0.357    & 0.346   \\
8       & 1.463    & 0.525    & 0.558    & 0.538   \\
9       & 2.030    & 0.846    & 0.808    & 0.774   \\
10      & 2.697    & 1.247    & 1.106    & 1.054   \\
11      & 3.468    & 1.727    & 1.455    & 1.378   \\
12      & 4.342    & 2.286    & 1.853    & 1.746   \\
13      & 5.321    & 2.922    & 2.302    & 2.157   \\
14      & 6.405    & 3.633    & 2.800    & 2.610   \\
15      & 7.596    & 4.418    & 3.349    & 3.104   \\
16      & 8.892    & 5.274    & 3.947    & 3.638   \\
17      & 10.29    & 6.200    & 4.594    & 4.212   \\
18      & 11.80    & 7.194    & 5.291    & 4.824   \\
19      & 13.41    & 8.252    & 6.036    & 5.474   \\
20      & 15.13    & 9.374    & 6.830    & 6.161   \\
\hline
\end{tabular}
\caption{
The total cross section $\sigma_{\nu d}^{L=0}$
(in units of $10^{-42}$ cm$^2$)
for the $\nu d$ reaction
leading to the final two-nucleon s-state.
For each of the four reactions in Eqs.(1) and (2),
$\sigma_{\nu d}^{L=0}$ calculated in EFT*
is shown as a function of the incident neutrino energy 
$E_\nu$ [MeV].
For the cutoff parameter, $\Lambda=$ 600 MeV 
has been used.}
\label{table;cross-section}
\end{center}
\end{table}

\begin{figure}
\parbox{0.45\textwidth}{\epsfig{file=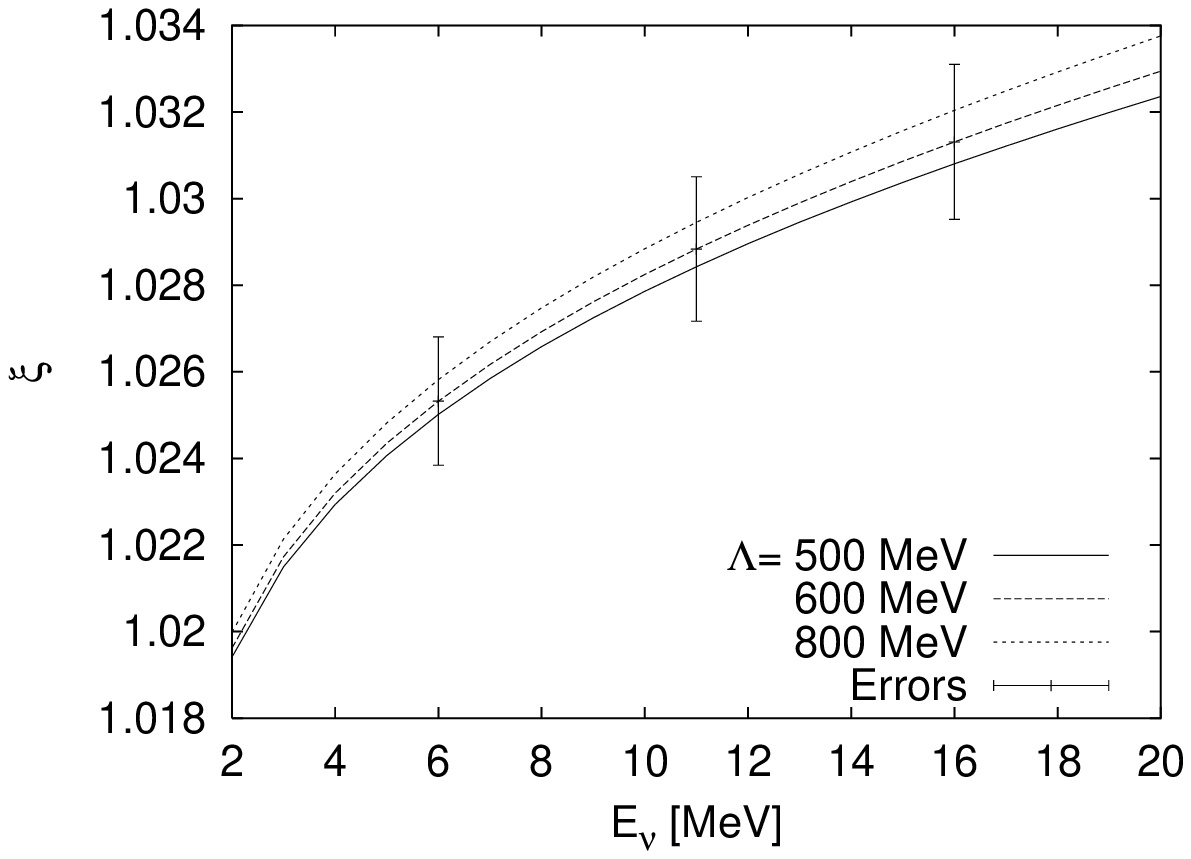,width=0.45\textwidth}
\caption{The ratio $\xi$ for CC defined in the text. 
The results for three different choices 
of $\Lambda$ are plotted.
The vertical bars represent changes 
in $\xi$ as $\hat{d}^R$ is varied within a range
allowed by the existing experimental errors in \G3H;
the representative results obtained 
for $\Lambda=600$ MeV are shown
for three values of $E_\nu$.}
\label{fig;2BCC}}
\hfill
\parbox{0.45\textwidth}{\epsfig{file=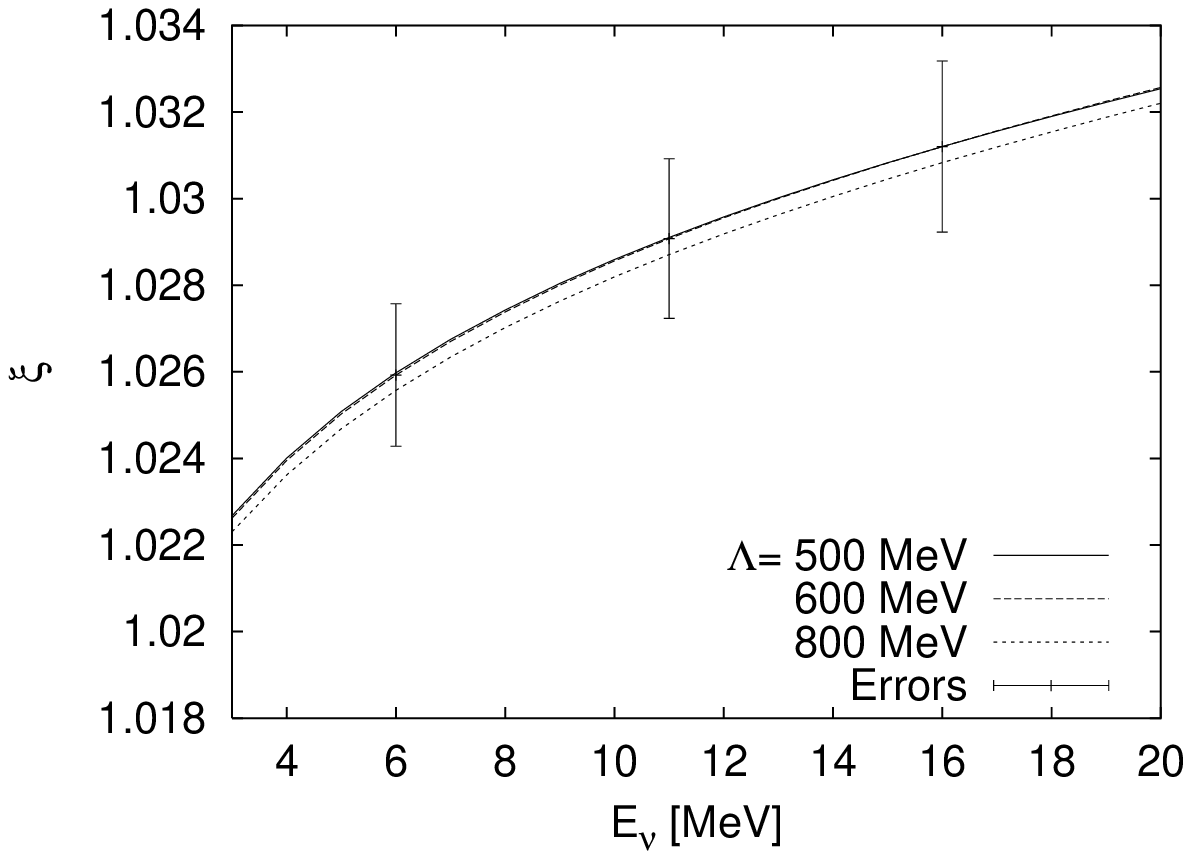,width=0.45\textwidth}
\caption{The ratio $\xi$ for NC defined in the text.
See also the caption for Fig. \ref{fig;2BCC}.
\mbox{ \ \ \ \ \ \ \ \ \ \ \ \ \ \ \ \ \ \ \ \ \ \ \ \ }
\mbox{ \ \ \ \ \ \ \ \ \ \ \ \ \ \ \ \ \ \ \ \ \ \ \ \ }
\mbox{ \ \ \ \ \ \ \ \ \ \ \ \ \ \ \ \ \ \ \ \ \ \ \ \ }
\mbox{ \ \ \ \ \ \ \ \ \ \ \ \ \ \ \ \ \ \ \ \ \ \ \ \ }
\mbox{ \ \ \ \ \ \ \ \ \ \ \ \ \ \ \ \ \ \ \ \ \ \ \ \ }
\mbox{ \ \ \ \ \ \ \ \ \ \ \ \ \ \ \ \ \ \ \ \ \ \ \ \ }
\mbox{ \ \ \ \ \ \ \ \ \ \ \ \ \ \ \ \ \ \ \ \ \ \ \ \ }
\mbox{ \ \ \ \ \ \ \ \ \ \ \ \ \ \ \ \ \ \ \ \ \ \ \ \ }
\mbox{ \ \ \ \ \ \ \ \ \ \ \ \ \ \ \ \ \ \ \ \ \ \ \ \ }
\mbox{ \ \ \ \ \ \ \ \ \ \ \ \ \ \ \ \ \ \ \ \ \ \ \ \ }
\mbox{ \ \ \ \ \ \ \ \ \ \ \ \ \ \ \ \ \ \ \ \ \ \ \ \ }
\mbox{ \ \ \ \ \ \ \ \ \ \ \ \ \ \ \ \ \ \ \ \ \ \ \ \ }
}
\label{fig;2BNC} }
\end{figure}

The results in Table \ref{table;cross-section} 
correspond to the case with $\Lambda=$ 600 MeV,
and we now discuss the cutoff dependence. 
In Figs. \ref{fig;2BCC} and \ref{fig;2BNC} 
we plot the ratio, 
$\xi \equiv \sigma_{1B+2B}/\sigma_{1B}$,
where $\sigma_{1B+2B}$ represents $\sigma_{\nu d}^{L=0}$
obtained with both the 1B and 2B currents included
while $\sigma_{1B}$ represents $\sigma_{\nu d}^{L=0}$
obtained with the 1B current alone.
Fig. \ref{fig;2BCC} gives $\xi$ for CC ($\nu d\to epp$),
while Fig. \ref{fig;2BNC} shows $\xi$
for NC ($\nu d\to \nu np$).
The three lines in each figure correspond
to different choices of $\Lambda$.
As can be seen from the figures,
$\sigma_{\nu d}^{L=0}$ 
exhibits extremely small $\Lambda$ dependence,
with only 0.02 \% changes over a wide range 
of physically reasonable values of $\Lambda$
($\Lambda=500 - 800$ MeV).

We now briefly discuss estimation 
of higher chiral-order effects.
The expansion parameter here is $Q/\Lambda$, 
where $Q$ is the pion mass $m_\pi$ 
or the typical external momentum scale $Q_{ext}$,
and $\Lambda$ is the chiral scale cutoff, 
$\Lambda\approx 600$ MeV.
It is common to assume $Q_{ext}\sim m_\pi$, 
but $Q_{ext}$ in our case is 
the incident neutrino energy $E_\nu$,
whose maximum value is $E_\nu^{max}\sim$20 MeV;
thus $E_\nu^{max}/\Lambda \simeq 0.03$
$\ll 0.23\simeq m_\pi/\Lambda$.
The actual numerical behavior 
of the chiral expansion in the present case
may be typified by the results for 
the CC reaction ($\nu d\to e^-pp$)
at $E_\nu = 20$ MeV. 
As far as the 1B operators are concerned,
the contribution to $\sigma^{L=0}$
of the LO terms amounts to 88.5 \%, 
while the corrections due to 
the NLO, N$^2$LO and N$^3$LO terms
are 8.8 \%, $-$0.5 \% and $\sim$ 0.001 \%,
respectively.
As for the 2B operators,
the N$^2$LO terms give a $\sim$ 0.3 \% correction,
whereas the N$^3$LO terms
give a $\sim 2.9$ \% correction.
Thus, the overall behavior is consistent
with convergence with respect to
the expansion parameter, $m_\pi/\Lambda$;
the rather conspicuous 2.9 \% correction
of the N$^3$LO 2B terms
is comparable to $(m_\pi/\Lambda)^3\simeq$ 1.2 \%,
while the other terms are decreasing faster
(almost in powers of $E_\nu^{max}/\Lambda$).
Therefore a possible measure of corrections 
due to N$^4$LO or higher-order terms
is 2.9 \%$\times (m_\pi/\Lambda)\sim$ 0.6 \%. 

The convergence property, however, 
can in fact be better than this.
Since in our approach
the overall strength, $\hat{d}^R$, of the 2B
operator is adjusted to reproduce $\Gamma^t_\beta$, 
the bulk of higher order corrections 
have already been effectively taken into account.
In particular, the chiral-symmetry breaking terms 
(proportional to $m_\pi$) give 
energy-independent contributions, 
which are essentially incorporated
into the effective $\hat{d}^R$.
The derivative terms acting on the wavefunctions 
or the two-body operators may pick up the pion mass scale, 
but their effects at the tritium $\beta$-decay energy  
are again essentially subsumed in $\hat{d}^R$. 
The remaining pieces of higher-order contributions
are $E_\nu$-dependent effects,
and hence they are likely to be controlled by 
the parameter $E_\nu/\Lambda$ rather than 
$m_\pi/\Lambda$. 
From this viewpoint it seems reasonable
to adopt 
2.9 \%$\times (E_\nu^{max}/\Lambda)\sim$ 0.1 \%
as a measure of the higher-order corrections.
Another measure of convergence is obtained 
as follows.
A tenet of a cutoff EFT
(such as used here) demands that,
provided an enough number of terms
are included in chiral expansion, 
the calculational results should be independent
of choices of the cutoff parameter $\Lambda$
(within a reasonable range).
Thus, the sensitivity of the calculated
$\sigma^{L=0}$ to $\Lambda$
serves as an indicator of the importance
of the contributions of 
the neglected higher order terms.
This sensitivity, however, has been found
to be extremely small (0.02 \% variation)
in our case.

Although the above discussion suggests
that higher-order effects (N$^4$LO or higher)
are reassuringly small,
we make a brief comment on 
three-body (3B) operators,
which represent 
a particular class of higher-order contributions.
It is known (see Table I of the last article
in Ref.~\cite{pphep}) that, at N$^4$LO, 
there is a contribution to the GT transition
from the 3B-operator, 
which we denote here by ${\cal {O}}_{GT}(3B)$.
Obviously, although ${\cal {O}}_{GT}(3B)$ 
contributes to \G3H, it plays no role
in the two-nucleon systems.
At N$^4$LO, therefore,
in renormalizing $\hat{d}^R$ with the use of \G3H,
one would need to subtract the contribution 
of ${\cal {O}}_{GT}(3B)$.
Formally speaking, our present treatment is
free from this complication,
since both the determination of $\hat{d}^R$
and the calculation of $\sigma_{\nu d}$
are carried out within N$^3$LO.
However, to the extent that 
$\hat{d}^R$ adjusted to reproduce
\G3H effectively includes higher order contributions,
the above-mentioned subtraction
is still needed.
Although a full solution of this problem
would require a systematic N$^4$LO calculation,
it is reasonable to expect that
the contributions of the 3B operators,
and hence the uncertainties due to them also,
lie within the above-discussed overall range 
of higher-order effects.

These considerations lead to the estimation
that the corrections due to the N$^4$LO 
or higher-order terms
should be of the order of $\sim$ 0.1 \%.
We also note that, within SNPA, the 3B contribution to \G3H
was calculated explicitly and found to be negligibly
small compared with the leading 2B terms 
\cite{3BC}.   

Figs. 2 and 3 also show
the uncertainty in $\xi$ due to the finite precision
with which $\hat{d}^R$ can be fixed from \G3H.
In fact, the largest uncertainty
in our present calculation comes from this origin,
and yet it only amounts to $\sim$ 0.5 \% ambiguity in $\xi$.
Based on these observations,
we consider it safe to conclude that $\sigma_{\nu d}^{L=0}$'s 
calculated here are reliable at the $\sim$ 1 \% level.

Comparison of our EFT* results 
with those of the latest SNPA calculation
by NETAL \cite{Netal} 
has already been described in \cite{Netal}.
We therefore only mention here
that $\sigma_{\nu d}^{L=0}$  
in table \ref{table;cross-section}
agrees with $\sigma_{\nu d}^{L=0}$ 
of NETAL within 1 \% accuracy 
(see Table 4 in \cite{Netal}).
As discussed, to the chiral order we are concerned with,
$\sigma_{\nu d}^{L\ge 1}$ calculated in EFT*
should agree with that obtained in SNPA.
Therefore $\sigma_{\nu d}$ (including all 
final partial waves) in EFT* 
can be identified, within 1 \% accuracy,
with $\sigma_{\nu d}$ given in NETAL\cite{Netal}.

There have been attempts 
to directly apply EFT to nuclear systems
with mass number A$\ge 3$ \cite{epel,kolck}.
Here, ``directly" means that the nuclear wave functions
are obtained in the framework of EFT
instead of using phenomenological potentials.
It will be interesting to employ
this ``direct" EFT approach 
for determining \dR\  (or $L_{1A}$) from \G3H
and use the resulting value of \dR\
for recalculating \signud.

To summarize, we have carried out an EFT* calculation
(up to N$^3$LO)
to estimate $\sigma_{\nu d}^{L=0}$,
the cross sections of the $\nu d$ reactions 
leading to the final two-nucleon $s$-wave state.
Our results agree, within 1 \% accuracy,
with those of the most recent SNPA calculation 
reported in \cite{Netal}.
In addition, we have found 
that the calculated $\sigma_{\nu d}^{L=0}$
exhibits very small cut-off dependence
(only $\sim$ 0.02 \% variation).
The corrections due to higher chiral order terms
are estimated to be of the order of $\sim$ 0.1 \%.
The prime uncertainties in the calculated 
$\sigma_{\nu d}^{L=0}$
stem from the experimental errors in \G3H;
this uncertainty, however, is less than $\sim$ 0.5 \%.
We therefore conclude that, 
apart from the radiative corrections
for which we refer to the literature,
the uncertainties in the calculated 
$\sigma_{\nu d}^{L=0}$
are less than 1 \%.

\vskip 3mm \noindent
{\bf Acknowledgments}

We are grateful to S. Nakamura and T. Sato 
for useful discussions. 
Thanks are also due to M. Rho for his interest 
in the present work and for his encouragement.
This work is supported in part by the US NSF,
Grant Nos.\ PHY-9900756 and INT-9730847,
by the BK21 Project of the Ministry of
Education of Korea, and 
by a grant from the Natural Sciences 
and Engineering Research Council of Canada.

\end{document}